\documentclass[10pt,journal,epsfig]{IEEEtran}
\usepackage{cite}
\usepackage{amsmath,amssymb,amsfonts}
\usepackage{algorithmic}
\usepackage{graphicx}
\usepackage{amssymb,comment}
\usepackage{algorithmic}
\usepackage{algorithm}
\usepackage{caption} 
\usepackage{subfig}
\usepackage{subcaption}
\usepackage{scrextend}
\usepackage{verbatim}
\usepackage{color,comment}
\usepackage{empheq}

\newtheorem{theorem}{\bf Theorem}
\newtheorem{lemma}{\bf Lemma}

\newcommand{\mbf}[1]{\mathbf{#1}}

\newcommand{\ic}[1]{\in\mathcal{#1}}

\usepackage{setspace}
\begin{document}
\title{\LARGE Movable-Antenna-Enhanced Physical-Layer Service Integration: Performance Analysis and Optimization}
\author{Xuanlin~Shen, Xin Wei,
        Weidong~Mei,~\IEEEmembership{Member,~IEEE,}
        %,~\IEEEmembership{Student Member,~IEEE,} 
        Zhi Chen,~\IEEEmembership{Senior Member,~IEEE,}
        Jun Fang,~\IEEEmembership{Senior Member,~IEEE,}
        Boyu Ning,~\IEEEmembership{Member,~IEEE}\vspace{-6pt}
\thanks{X. Shen is with the Glasgow College, University of Electronic Science and Technology of China, Chengdu, China (e-mail: xlshen@std.uestc.edu.cn).}
\thanks{X. Wei, W. Mei, Z. Chen, J. Fang, and B. Ning are with the National Key Laboratory of Wireless Communications, University of Electronic Science and Technology of China, Chengdu, China (e-mail: xinwei@std.uestc.edu.cn, wmei@uestc.edu.cn, chenzhi@uestc.edu.cn, JunFang@uestc.edu.cn, boydning@outlook.com).}}
\maketitle

\begin{abstract}
Movable antennas (MAs) have drawn increasing attention in wireless communications due to their capability to create favorable channel conditions via local movement within a confined region. In this letter, we investigate its application in physical-layer service integration (PHY-SI), where a multi-MA base station (BS) simultaneously transmits both confidential and multicast messages to two users. The multicast message is intended for both users, while the confidential message is intended only for one user and must remain perfectly secure from the other. Our goal is to jointly optimize the secrecy and multicast beamforming, as well as the MAs' positions at the BS to maximize the secrecy rate for one user while satisfying the multicast rate requirement for both users. To gain insights, we first conduct performance analysis of this MA-enhanced PHY-SI system in two special cases, revealing its unique characteristics compared to conventional PHY-SI with fixed-position antennas (FPAs). To address the secrecy rate maximization problem, we propose a two-layer optimization framework that integrates the semidefinite relaxation (SDR) technique and a discrete sampling algorithm. Numerical results demonstrate that MAs can greatly enhance the achievable secrecy rate region for PHY-SI compared to FPAs.
\end{abstract}
\begin{IEEEkeywords}
Movable antennas, physical-layer service integration, secrecy rate region, performance analysis, discrete sampling.
\end{IEEEkeywords}\vspace{-6pt}
\IEEEpeerreviewmaketitle

\section{Introduction}
Physical-layer service integration (PHY-SI) refers to the simultaneous transmission of public (multicast) and confidential (secrecy) information at the physical layer using superposition coding \cite{R. Schaefer}. This technique is expected to meet the dual demands of high data rates and strong security in future wireless communication systems \cite{R. Schaefer, W. Mei, B. Ning}. However, conventional PHY-SI relies on fixed-position antennas (FPAs), which can limit the available design degrees of freedom (DoFs) for both secure and multicast signal transmission.

Recently, movable antenna (MA) technology has received increasing attention in the field of wireless communications. Compared to conventional FPAs, MAs allow for flexible antenna movement within a confined region at the transmitter/receiver side to create favorable channel conditions and enhance antenna aperture, thereby dramatically improving the wireless communication and sensing performance \cite{L. Zhu2024,L. Zhu2025tutorial}. Inspired by the promising benefits of MAs, some existing works have delved into their position optimization under various scenarios, e.g., array signal processing \cite{L. Zhu2023, D. Wang},  single-input single-output (SISO) \cite{X. Zeng}, multi-user MISO \cite{L. ZhuMAC}, cognitive radio \cite{X. Wei}, among others. Particularly, the authors in \cite{Y. Gao} and \cite{G. Hu}, \cite{Z. Cheng} have shown the significant performance advantages of MAs over FPAs in terms of secrecy and multicast transmission, respectively. However, to the best of our knowledge, there is no existing work focusing on the application of MAs in PHY-SI, where the MA position optimization needs to cater to both secrecy and multicast transmission, as well as their mutual interference.

\begin{figure}[!t]
\centering
\includegraphics[width=0.38\textwidth]{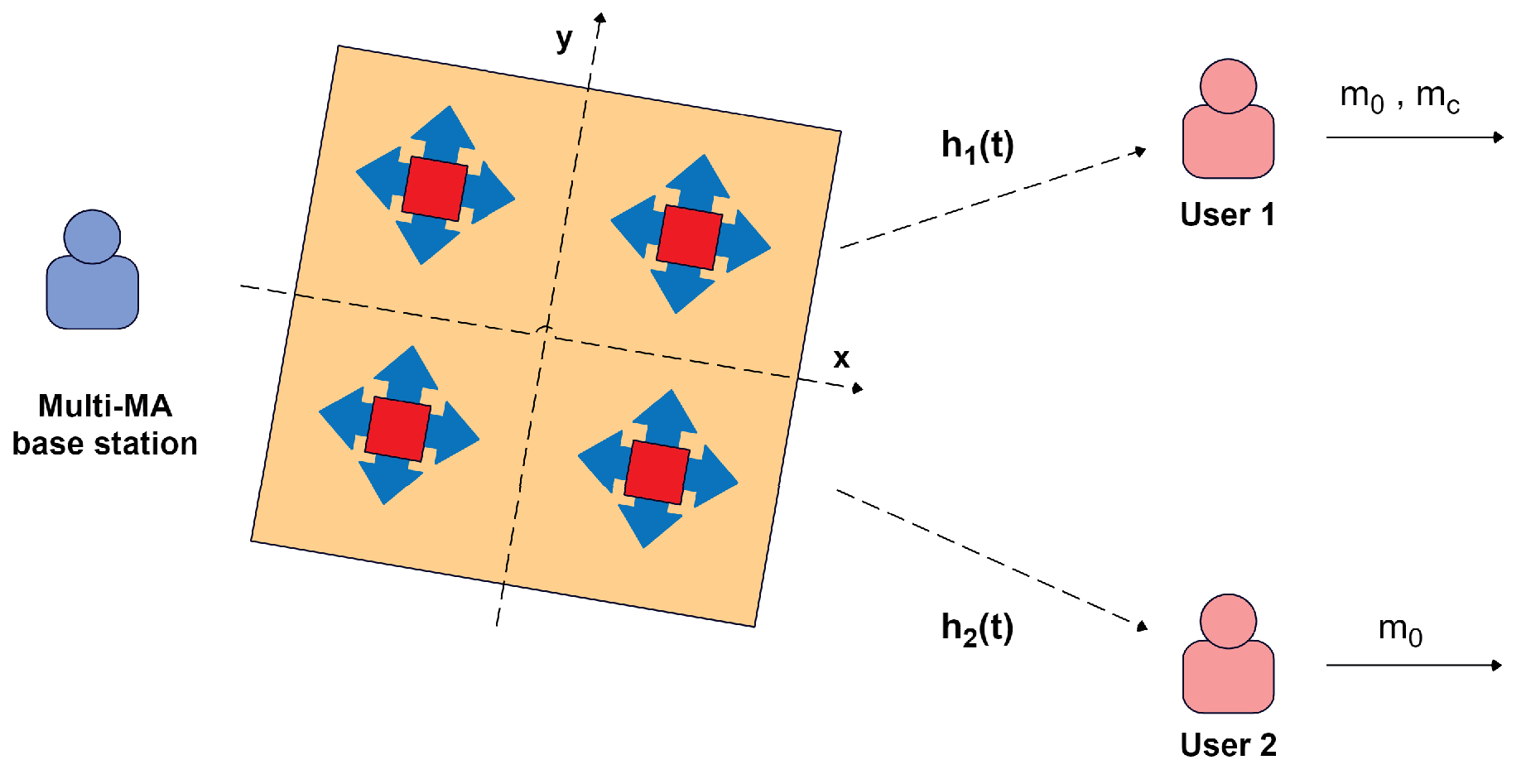}
\caption{MA-enhanced MISO PHY-SI system.}
\label{Fig_SysModel}
\vspace{-15pt}
\end{figure}

To fill in this gap, in this letter, we investigate an MA-enhanced PHY-SI system as shown in Fig.\,\ref{Fig_SysModel}, where a multi-MA base station (BS) simultaneously transmits both confidential and multicast messages to two users. The multicast message is intended for both users, while the confidential message is intended only for one user and must be kept perfectly secure from the other. We aim to jointly optimize the secrecy and multicast beamforming, as well as the MAs' positions at the BS to maximize the secrecy rate at one user while satisfying the constraint on the multicast rate for both users. However, this secrecy rate maximization (SRM) problem is difficult to be optimally solved due to the coupling of the two-type signal transmission. To obtain insights, we first conduct performance analysis of this MA-enhanced PHY-SI system in two special cases. It is shown that PHY-SI with a single MA may be outperformed by a time-sharing strategy, while employing multiple MAs can restore its advantages. To solve the SRM problem, we propose a two-layer optimization framework that integrates the semi-definite relaxation (SDR) technique and the discrete sampling algorithm. Numerical results demonstrate that the proposed algorithm can greatly enhance the achievable secrecy rate region for PHY-SI compared to FPAs and particle swarm optimization (PSO) algorithm.\vspace{-6pt}

\begingroup
\allowdisplaybreaks
\section{System Model and Problem Formulation}
As shown in Fig.\,\ref{Fig_SysModel}, we consider an MA-enhanced PHY-SI system, with a multi-MA BS and two users. It is assumed that both users subscribe to the multicast service and user 1 also subscribes to the confidential service. Hence, user 2 is regarded as a potential
eavesdropper as far as the confidential message for user 1 is concerned. To enhance the performance of the PHY-SI system, the BS is equipped with $N$ MAs that can be flexibly moved within a two-dimensional (2D) region, which is a square of $A \times A$ and denoted as ${\cal C}_t$. Without loss of generality, we establish a coordinate system and assume that $\mathcal{C}_t$ is parallel to the $x$-$y$ plane, as shown in Fig. 1. Denote by $\mathbf{t}_n=[x_n,y_n]^T\in\mathcal{C}_t, n \in {\cal N}\triangleq\{1,2,\cdots,N\}$, the position of the $n$-th transmit MA in $\mathcal{C}_t$, given a reference origin point $\mathbf{t}_0= [0, 0]^T$. We assume that the antenna movement delay is much smaller than channel coherence time, which usually holds in various slow-varying scenarios, e.g., smart homes/factories.

By stacking the $N$ MAs' positions at the BS, we define its antenna position vector (APV) as $\mathbf{T}=[\mathbf{t}_1,\mathbf{t}_2,\cdots,\mathbf{t}_N]\in\mathbb{C}^{2\times N}$ and denote its downlink channels with user $k$ as $\mathbf{h}^H_k(\mathbf{T}) \in \mathbb{C}^{1\times N}$, $ k = 1, 2$. 
Therefore, the received signal at user $k$ is given by
\begin{equation}
    y_k=\mathbf{h}^H_k(\mathbf{T}) \mathbf{x} + z_k, k=1, 2,
\end{equation}
where $z_k$'s are independent identically distributed (i.i.d.) complex Gaussian noise with zero mean and variance of $\sigma^2$, and $\mathbf{x} \in \mathbb{C}^{N\times 1}$ is the transmitted signal vector consisting of two independent components, i.e., $\mathbf{x} = \mathbf{w}_0m_0 + \mathbf{w}_cm_c$. Here, $m_0$ is the multicast message intended for both users, $m_c$ is the confidential message intended for user 1 only, and $\mathbf{w}_0\in\mathbb{C}^{N\times 1}$ and $\mathbf{w}_c\in\mathbb{C}^{N\times 1}$ denote the multicast and secrecy beamforming vectors, respectively.

Denote $R_0(\mathbf{T})$ and $R_c(\mathbf{T})$ as the achievable multicast and secrecy rates, respectively. Then, an achievable secrecy rate region of the considered PHY-SI system for any given APV $\mathbf{T}$ can be characterized as the set of all nonnegative rate pairs $(R_0(\mathbf{T}),R_c(\mathbf{T}))$ satisfying\cite{R. Schaefer}
\begingroup\makeatletter\def\f@size{9}\check@mathfonts
\def\maketag@@@#1{\hbox{\m@th\normalsize\normalfont#1}}%
\begin{align}
        R_0(\mathbf{T}) &\leq \min_{k=1, 2} \log_2\left(1 + \frac{|\mathbf{h}_k^H\left( \mathbf{T} \right) \mathbf{w}_0|^2 }{\sigma^2 + |\mathbf{h}_k^H\left( \mathbf{T} \right) \mathbf{w}_c|^2 }\right), \nonumber\\
        R_c(\mathbf{T}) &\leq \log_2\left(1\!+\!\frac{|\mathbf{h}_1^H (\mathbf{T})\mathbf{w}_c |^2}{\sigma^2}\right)\!-\!\log_2 \left( 1\!+\!\frac{|\mathbf{h}_2^H (\mathbf{T})\mathbf{w}_c |^2}{\sigma^2}\right) \nonumber\\ 
        &=\log_2\left(\frac{\sigma^2 + |\mathbf{h}_1^H (\mathbf{T})\mathbf{w}_c |^2}{\sigma^2 + |\mathbf{h}_2^H (\mathbf{T})\mathbf{w}_c |^2}\right).\label{eqn_ratepairs}
\end{align}
\endgroup
In this letter, we aim to jointly optimize the multicast and secrecy beamforming vectors (i.e., $\mathbf{w}_0$ and $\mathbf{w}_c$), as well as the APV (i.e., $\mathbf{T}$) to maximize the secrecy rate subject to the constraint on the multicast rate, i.e.,
\begingroup\makeatletter\def\f@size{9}\check@mathfonts
\def\maketag@@@#1{\hbox{\m@th\normalsize\normalfont#1}}%
\begin{subequations}\label{eqn_SRM}
    \begin{align}
       \text{(P1)} & \quad \max_{\mathbf{w}_0, \mathbf{w}_c, \mathbf{T}}\log_2\left(\frac{\sigma^2 + |\mathbf{h}_1^H (\mathbf{T})\mathbf{w}_c |^2}{\sigma^2 + |\mathbf{h}_2^H (\mathbf{T})\mathbf{w}_c |^2}\right) \nonumber\\
        \text{s.t.} &\quad \min_{k=1, 2} \log_2\left(1 + \frac{|\mathbf{h}_k^H\left( \mathbf{T} \right) \mathbf{w}_0|^2 }{\sigma^2 + |\mathbf{h}_k^H\left( \mathbf{T} \right) \mathbf{w}_c |^2}\right) \geq r_{ms}, \\
        & \quad |\mathbf{w}_0|^2 + |\mathbf{w}_c|^2 \leq P, \\
        & \quad \boldsymbol{\mathrm{t}}_n \in \mathcal{C}_t, \quad n \in \mathcal{N}, \label{p1c}\\
        & \quad \|\boldsymbol{\mathrm{t}}_k - \boldsymbol{\mathrm{t}}_m\|_2 \geq D_{\min}, \quad k,m \in \mathcal{N}, \, k \neq m,\label{p1d}
    \end{align}
\end{subequations}\endgroup
where $P$ denotes the BS's maximum transmit power, $r_{ms}$ denotes a prescribed requirement on the multicast rate, and $D_{min}$ denotes the minimum spacing between any two MAs to avoid mutual coupling. All required channel information is assumed to be available by applying the techniques presented in \cite{L. Zhu2025tutorial} dedicated to MAs to characterize the performance limit.
\begin{comment}
Note that for (P1), if we set $r_{ms}=0$, it becomes an SRM problem in secure MA systems. On the contrary, the secrecy rate will become zero if $r_{ms}$ is set above a threshold given by
\begin{equation}\label{eqn_multicastcapacity}
    r_{\max}=\max_{|\mathbf{w}_0|^2\leq P}  \min_{k\in {1,2}} \log_2\left(1+\frac{|\mathbf{h}_k^H\left( \mathbf{T} \right) \mathbf{w}_0|^2}{\sigma^2}\right).
\end{equation}
Obviously, $r_{\max}$ is the maximum multicast rate for a two-user MA system.
\end{comment}

However, it is noted that (P1) is difficult to be optimally solved due to the intricate coupling between the transmit beamforming vector $\mathbf{w}$ and the APV $\mathbf{T}$. In the next section, we first analyze the achievable secrecy rate region of the MA-enhanced PHY-SI system under some special cases.\vspace{-8pt}

\section{Performance Analysis}
In this section, we conduct performance analysis of the MA-enhanced PHY-SI in two special cases with a single MA and line-of-sight (LoS) channels, respectively, to obtain insights.  \vspace{-9pt}

\subsection{Single-MA Case}\label{singleMACase}
First, we consider a single-MA case, in which the APV reduces to a vector $\mathbf{t} \in {\mathbb R}^{2 \times 1}$, and the MISO channel vectors reduces to scalars $h_k(\mathbf{t}), k=1,2$. Let $P_0$ and $P_c$ denote the power allocated to the multicast and confidential messages, respectively, with $P_0+P_c \leq P$. Then, (P1) reduces to
\begingroup\makeatletter\def\f@size{9}\check@mathfonts
\def\maketag@@@#1{\hbox{\m@th\normalsize\normalfont#1}}%
\begin{subequations}\label{eqn_P1_SISO}
    \begin{align}
       \text{(P1-SISO)} & \quad \max_{P_0, P_c, \mathbf{t}\in \mathcal{C}_t}\log_2\left(\frac{\sigma^2 + P_c|h_1 (\mathbf{t})|^2 }{\sigma^2 + P_c|h_2 (\mathbf{t})|^2 }\right) \nonumber\\
       \text{s.t.} \quad & \min_{k=1, 2} \log_2\left(1 + \frac{P_0|h_k\left( \mathbf{t} \right)|^2 }{\sigma^2 + P_c|h_k(\mathbf{t}) |^2}\right) \geq r_{ms}, \label{P1-SISOa}\\
        & \quad P_0 + P_c \leq P.\label{P1-SISOb}
    \end{align}
\end{subequations}\endgroup
For (P1-SISO), it is not difficult to see the equality in constraint (\ref{P1-SISOa}) and (\ref{P1-SISOb}) should hold at the optimality. Moreover, at the optimality of (P1-SISO), we should have $\lvert h_1(\mathbf{t}) \rvert^2 \ge \lvert h_2(\mathbf{t}) \rvert^2$ to achieve a non-negative secrecy rate. Hence, the minimum at the left-hand side of (\ref{P1-SISOa}) should be achieved at $k=2$. Based on these observations, it can be shown that the optimal power allocation is given by
\begin{equation}\label{eqn_power}
    P_0^*=\frac{\tau_{ms}(P|h_{2}(\mathbf{t})|^2+\sigma^2)}{(\tau_{ms}+1)|h_2(\mathbf{t})|^2},\;\;P_c^*=P-P_0^*,
\end{equation}
where $\tau_{ms}=2^{r_{ms}}-1$. 

By substituting \eqref{eqn_power} into the objective function of (P1-SISO), the maximum secrecy rate for any given $\mathbf{t}$ can be obtained as
\begin{equation}
    R_c^*(\mathbf{t})=\log_2\left(\frac{A(\mathbf{t}) + (P|h_2(\mathbf{t})|^2-\tau_{ms}\sigma^2)|h_1(\mathbf{t})|^2}{(\sigma^2 + P|h_2(\mathbf{t})|^2)|h_2(\mathbf{t})|^2}\right), 
\end{equation}
where $A(\mathbf{t})=(\tau_{ms}+1)\sigma^2 |h_2(\mathbf{t})|^2$.

Then, the optimal MA position can be obtained as
$\mathbf{t}^*=\arg \max_{\mathbf{t} \in \mathcal{C}_t} R_c^*(\mathbf{t})$,
which can be approximately optimally solved by performing an exhaustive search within $\mathcal{C}_t$ by sampling it into a multitude of discrete points.

Given the optimal solution to (P1-SISO), we next compare the secrecy rate region of the single-MA PHY-SI system with that of time-sharing strategy, where the multicast and confidential messages are transmitted to the two users within two orthogonal time slots, respectively. The associated secrecy rate region is given by the union of the following rate pairs,
\begin{equation}
    \begin{cases}
     R_c^\text{TS}(\mathbf{t})=\alpha \log_2\left(\frac{\sigma^2 + P|h_1(\mathbf{t})|^2}{\sigma^2 + P|h_2 (\mathbf{t})|^2}\right),\\
    R_{0}^\text{TS}(\mathbf{t})=(1-\alpha) \log_2\left(1 + \frac{P|h_2(\mathbf{t})|^2 }{\sigma^2}\right),
    \end{cases}
\end{equation} 
for all $\alpha \in [0,1]$. Note that for conventional PHY-SI system with a fixed FPA, denoted as ${\mathbf{t}}_{\text{FPA}}$,  PHY-SI always gives rise to a larger rate region compared to time sharing. This is because the secrecy and multicast rates are both concave in the power allocation factor. Hence, we have $R_c^\text{TS}(\mathbf{t}_\text{FPA}) \le \log_2\left(\frac{\sigma^2 + P\alpha|h_1({\mathbf{t}}_{\text{FPA}})|^2}{\sigma^2 + P\alpha|h_2 ({\mathbf{t}}_{\text{FPA}})|^2}\right)$
and $R_{0}^\text{TS}(\mathbf{t}_\text{FPA}) \le \log_2\left(1 + \frac{P(1-\alpha)|h_2(\mathbf{{\mathbf{t}}_{\text{FPA}}})|^2 }{\sigma^2}\right)$.

However, we argue that the superiority of PHY-SI over time-sharing may not hold in single-MA systems. The key reason is that the optimal MA positions for maximizing the secrecy rate and multicast rate, respectively, denoted as $\mathbf{t}_c$ and $\mathbf{t}_0$, may differ. Since the PHY-SI employs a common MA position for concurrent multicast and secrecy transmission, no MA position may exist to yield a better rate pair than time sharing element-wise. Note that this MA position cannot be found via ``position-sharing'' between $\mathbf{t}_c$ and $\mathbf{t}_0$, as the secrecy rate and multicast rate are generally highly nonlinear (and thus non-concave) functions of $\mathbf{t}$.

To verify this claim, we show the achievable secrecy rate regions of the single-MA PHY-SI system under two channel setups in Figs.\,\ref{singleMA}(a) and \ref{singleMA}(b), respectively. For these two figures, we employ the same field-response channel model as in Section V, with the same angles of departure (AoDs) from the BS to the two users but different randomly selected path gains for the transmit paths.  It is observed that the secrecy rate region with an MA is larger than that with an FPA under both setups. However, the time-sharing strategy can outperform PHY-SI under the setup in Fig.\,\ref{singleMA}(a), which validates our claim. However, unlike the single-MA case presented in this subsection, employing multiple MAs may help retrieve the performance gain of PHY-SI over time sharing thanks to the joint beamforming and APV design, as shown next.\vspace{-9pt}
\begin{figure}[t]
    \centering
    \begin{minipage}[]{0.47\columnwidth}
        \includegraphics[width=\linewidth]{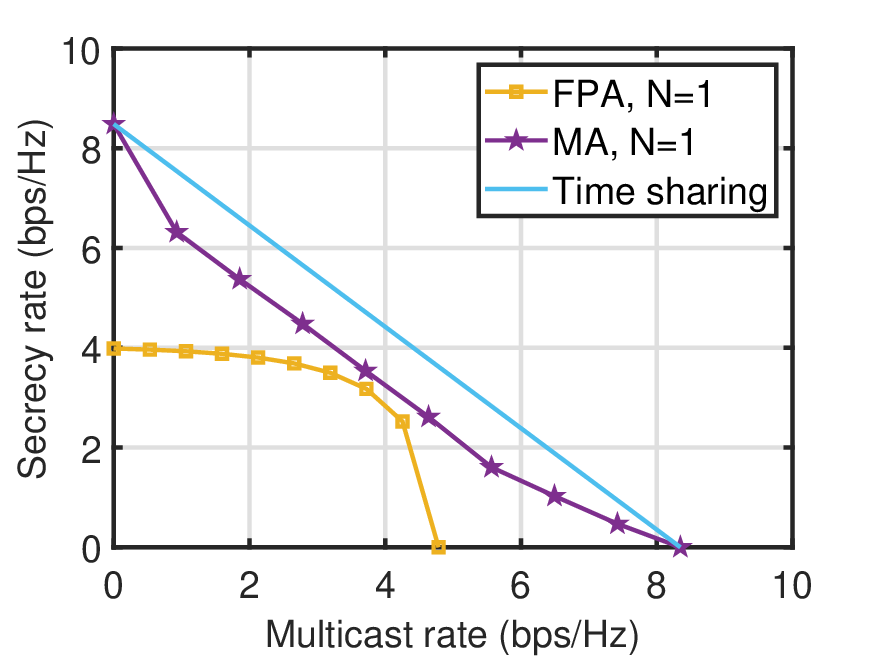}
        \caption*{(a)}
    \end{minipage}\hspace{0.001\columnwidth}% <-- 关键控制点
    \begin{minipage}[]{0.47\columnwidth}
        \includegraphics[width=\linewidth]{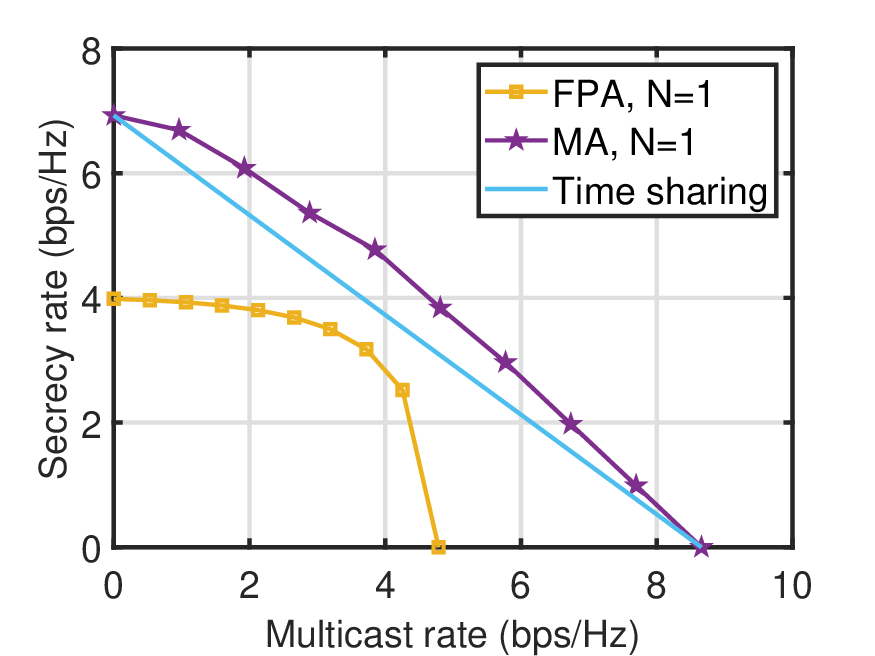}
        \caption*{(b)}
    \end{minipage}
    \vspace{-0.5em}
    \caption{Comparison between PHY-SI and time sharing for a single-MA system}\vspace{-2em}
    \label{singleMA}
\end{figure}

\subsection{Special Case with LoS Channels}    
In this subsection, we consider another special case with LoS channels from the BS to both users. Let $\theta_{k}$/$\phi_{k}$ be the elevation/azimuth AoD from the BS to user $k, k=1,2$. Then, the array response from the BS to user $k$ can be written as
\begin{equation}\label{eqn_ArrayResp}
    \mathbf{a}\left( \mathbf{T}, \theta_{k},\phi_{k} \right) = [e^{j\rho _{k}\left( \mathbf{t}_1 \right)},e^{j\rho _{k}\left( \mathbf{t}_2 \right)},
    \cdots ,e^{j\rho _{k}\left( \mathbf{t}_N \right)}],
\end{equation}
where $\lambda$ denotes the wavelength, and $\rho_{k}\left( \boldsymbol{\mathrm{t}}_n \right) =\frac{2\pi}{\lambda}(x_n\sin \theta _{k}\cos \phi _{k}+y_n\cos \theta _{k})$. As such, the channel vector from the BS to user $k$ can be expressed as $\mbf{h}_k(\mbf{T})=\beta_k\mbf{a}(\mbf{T},\theta_k,\phi_k),\,\,k=1,2$,
where $\beta_{k}$ denotes the path gain from the BS to user $k$.

Then, the achievable secrecy rate and multicast rate can be expressed as
\begingroup\makeatletter\def\f@size{9}\check@mathfonts
\def\maketag@@@#1{\hbox{\m@th\normalsize\normalfont#1}}%
\begin{align}	&R_c(\mbf{T})=\log_2\frac{\sigma^2+P_c\left|\beta_1\right|^2\left|\mbf{w}_c^H\mbf{a}(\mbf{T},\theta_1,\phi_1)\right|^2}{\sigma^2+P_c\left|\beta_2\right|^2\left|\mbf{w}_c^H\mbf{a}(\mbf{T},\theta_2,\phi_2)\right|^2},\label{eqn_LoS_Rc}\\
&R_0(\mbf{T})=\underset{k=1,2}{\min}R_{0,k}(\mbf{T})\label{eqn_LoS_R0},
\end{align}\endgroup
where $R_{0,k}(\mbf{T})=\log_2\left(1+\frac{P_0\left|\beta_k\right|^2\left|\mbf{w}_0^H\mbf{a}(\mbf{T},\theta_k,\phi_k)\right|^2}{\sigma^2+P_c\left|\beta_k\right|^2\left|\mbf{w}_c^H\mbf{a}(\mbf{T},\theta_k,\phi_k)\right|^2}\right)$.  We further assume an arbitrarily large movement region as in \cite{L. Zhu2023}, \cite{X. Wei}, \cite{L. Zhu2024Modeling} (i.e., constraint \eqref{p1c} is relaxed).\footnote{It is worth noting that the actual movement region of the antennas is always finite. The purpose of the asymptotic analysis here is to reveal the unique characteristics of MAs compared to FPAs in PHY-SI, similarly to that for massive multiple-input multiple-output (MIMO) systems \cite{MIMO}.} Then, the optimal secrecy and multicast beamforming solutions to (P1) can always be obtained as array response vectors, as presented in the following lemma and theorem. 	
\begin{lemma}
Let $\tilde{\mbf{w}}_c(\mbf{T})=\frac{1}{\sqrt{N}}\mbf{a}(\mbf{T},\theta_c,\phi_c)$ and $\tilde{\mbf{w}}_0(\mbf{T})=\frac{1}{\sqrt{N}}\mbf{a}(\mbf{T},\theta_0,\phi_0)$ denote the BS's transmit beamforming vector for the confidential and multicast messages, respectively, where $\theta_{c}$/$\phi_{c}$ and $\theta_{0}$/$\phi_{0}$ denote their elevation/azimuth AoDs, which can be arbitrarily selected as long as $\theta_0\ne\theta_c\ne\theta_k, k=1, 2$. Define 
\begin{equation}\label{eqn_Lem1_func}
g_{i,k}(\mbf{T})\triangleq\left|\tilde{\mbf{w}}_i^H(\mbf{T})\mbf{a}(\mbf{T},\theta_k,\phi_k)\right|^2\!,i\in\left\{c,0\right\}\!,\!k=1,2,
\end{equation}
as the secrecy/multicast beamforming gain at the direction of user $k$.
Then, there exists an APV $\mbf{T}^{\star}$ such that
\begin{equation}\label{null}
g_{c,1}(\mbf{T}^{\star})=g_{0,1}(\mbf{T}^{\star})=g_{0,2}(\mbf{T}^{\star})=N,\; g_{c,2}(\mbf{T}^{\star})=0.
\end{equation}
\end{lemma}
\begin{IEEEproof}
To construct $\mbf{T}^{\star}$, we consider that all MAs are placed with an equal spacing (denoted as $d$) along the horizontal dimension. Then, the position of the $n$-th MA is given by $\mbf{t}_n=\left[(n-1)d,0\right]^T$, $n\ic{N}$, and $g_{i,k}(\mbf{T})$ in \eqref{eqn_Lem1_func} can be simplified as
%\begingroup\makeatletter\def\f@size{9}\check@mathfonts\def\maketag@@@#1{\hbox{\m@th\normalsize\normalfont#1}}%
\begin{equation}\label{eqn_Lem1_eqn1}
g_{i,k}(\mbf{T})=\frac{1}{N}\left|\sum_{n=1}^{N}e^{j2\pi(n-1)dr_{i,k}}\right|^2=\frac{1}{N}\frac{1-\cos \left(2\pi Nr_{i,k}d\right)}{1-\cos \left(2\pi r_{i,k}d\right)},
\end{equation}
%\endgroup
where $r_{i,k}\triangleq\frac{1}{\lambda}\left(\sin\theta_k\cos\phi_k-\sin\theta_i\cos\phi_i\!+\!\cos\theta_k\!-\!\cos\theta_i\right)$.
		
Note that due to the randomly distributed scatterers in the environment, all $r_{i,k}$'s can be considered as irrational values and independent over the set of rational numbers with the probability of one. As such, according to \cite{A. Leshem}, for any given $\epsilon\in(0,2\pi)$, there must exist a positive integer $d$ such that
\begin{align}
    &0 < 2\pi Nr_{i,1}d < \epsilon, \;0 < 2\pi r_{i,1}d < \epsilon\;(\mathrm{mod}\,\,2\pi),\; i\in\left\{c,0\right\},\nonumber\\
    &0 < 2\pi Nr_{0,2}d < \epsilon,\;0 < 2\pi r_{0,2}d < \epsilon\;(\mathrm{mod}\,\,2\pi),\nonumber\\
    &0 < 2\pi Nr_{c,2}d < \epsilon\;(\mathrm{mod}\,\,2\pi)\nonumber\\
    &\pi - \frac{\epsilon}{2} < 2\pi r_{c,2}d < \pi + \frac{\epsilon}{2}\;(\mathrm{mod}\,\,2\pi).\label{eqn_Lem1_eqn2}
\end{align}
% \begin{equation}\label{eqn_Lem1_eqn2}
% \begin{cases}
%     \cos \left(2\pi Nr_{c,1}d\right) >  1-\delta,\quad\cos \left(2\pi r_{c,1}d\right) >  1-\delta,\\
%     \cos \left(2\pi Nr_{c,2}d\right) >  1-\delta,\quad\cos \left(2\pi r_{c,2}d\right) <  \delta,\\
%     \cos \left(2\pi Nr_{0,1}d\right) >  1-\delta,\quad\cos \left(2\pi r_{0,1}d\right) >  1-\delta,\\
%     \cos \left(2\pi Nr_{0,2}d\right) >  1-\delta,\quad\cos \left(2\pi r_{0,2}d\right) >  1-\delta.\\
% \end{cases}
% \end{equation}
By substituting \eqref{eqn_Lem1_eqn2} into \eqref{eqn_Lem1_eqn1} and letting $\epsilon\rightarrow0$, we can obtain \eqref{null}.
%\begin{equation}
    %\begin{cases}
        %\underset{r_{i,1}d\rightarrow0}{\lim}g_{i,1}(\mbf{T})=\underset{r_{i,1}d\rightarrow0}{\lim}g_{0,2}(\mbf{T})= N\\
        %\underset{2\pi Nr_{c,2}d\rightarrow0,\,\,2\pi r_{c,2}d\rightarrow\pi}{\lim}g_{c,2}(\mbf{T})=0
    %\end{cases}
%\end{equation}
%\begin{equation}
%g_{c,1}(\mbf{T})\rightarrow N,\,\,g_{0,1}(\mbf{T})\rightarrow N,\,\, g_{0,2}(\mbf{T})\rightarrow N,\,\, g_{c,2}(\mbf{T})\rightarrow 0.\nonumber
%\end{equation}
Moreover, as $d$ is a positive integer that is much greater than the wavelength-level antenna spacing in practice, the minimum antenna spacing constraints in \eqref{eqn_SRM} can be satisfied. This completes the proof.
\end{IEEEproof}

The equalities in \eqref{null} reveal that via MA position optimization, the confidential beam gain can be maximized at user 1 while being nulled at user 2. Moreover, the multicast beam gain can be maximized at both users 1 and 2. With Lemma 1, we proceed to show the optimality of $\mathbf{T}^{\star}$.
\begin{theorem}
The APV $\mathbf{T}^{\star}$ can result in a larger secrecy rate region achievable by any other APVs $\mbf{T}$, i.e., $R_0(\mbf{T}^{\star}) \ge R_0(\mathbf{T})$ and $R_c(\mbf{T}^{\star})\ge R_c(\mathbf{T})$.
\end{theorem}
\begin{IEEEproof}
For the APV $\mathbf{T}^{\star}$ and its associated beamforming designs $\tilde{\mbf{w}}_c(\mbf{T})$ and $\tilde{\mbf{w}}_0(\mbf{T})$ presented in Lemma 1, the achievable secrecy rate and multicast rate are given by
\begin{align}
&R_c(\mbf{T}^{\star})=\log_2\left(1+\frac{\tilde{P}_cN\left|\beta_1\right|^2}{\sigma^2}\right),\label{eqn_Th1_Rc}\\
&R_0(\mbf{T}^{\star})=\underset{k=1,2}{\min}R_{0,k}(\mbf{T}^{\star})\label{eqn_Th1_R0},
\end{align}
where $R_{0,1}(\mbf{T})=\log_2\left(1+\frac{\tilde{P}_0N\left|\beta_1\right|^2}{\sigma^2+\tilde{P}_cN\left|\beta_1\right|^2}\right),\label{eqn_Th1_R01}$
    and $R_{0,2}(\mbf{T})=\log_2\left(1+\frac{\tilde{P}_0N\left|\beta_2\right|^2}{\sigma^2}\right)\label{eqn_Th1_R02}$, and $\tilde{P}_c$ and $\tilde{P}_0$ denote the power allocation to the confidential and multicast messages, respectively. Then, by comparing \eqref{eqn_LoS_Rc}-\eqref{eqn_LoS_R0} with \eqref{eqn_Th1_Rc}-\eqref{eqn_Th1_R0}, if we set       
\begin{equation}\label{eqn_Th1_TxPw}
    \tilde{P}_c=P_c\frac{\left|\mbf{w}_c^H\mbf{a}(\mbf{T},\theta_1,\phi_1)\right|^2}{N},\quad \tilde{P}_0=P-\tilde{P}_c,
\end{equation}
and substitute \eqref{eqn_Th1_TxPw} into \eqref{eqn_Th1_Rc}-\eqref{eqn_Th1_R0}, we can always obtain $R_0(\mbf{T}^{\star}) \ge R_0(\mathbf{T}) $ and  $R_c(\mbf{T}^{\star})\ge R_c(\mathbf{T})$. 
\end{IEEEproof}

Notably, since the APV solution presented in Lemma 1 can yield a larger secrecy rate region compared to any other APVs, it should also outperform the time-sharing strategy. Theorem 1 also manifests the exceptional capability of MAs to configure signal transmission of different types, beyond the single-type signal transmission as studied in \cite{Y. Gao, G. Hu, Z. Cheng}.\vspace{-6pt}
	
\section{Proposed Solution to (P1)}
In this section, we aim to solve (P1) in the general case. To this end, we propose a two-layer optimization algorithm to decompose (P1) into two subproblems. The inner problem optimizes $\mathbf{w}_0$ and $\mathbf{w}_c$ for a given APV $\mathbf{T}$, while the outer problem optimizes the APV based on the optimal value of the inner problem. The outer and inner optimization problems are respectively given by
\begin{equation}
   \text{(P2-Outer)}\; \max_\mathbf{T}\; g^*(\mathbf{T}),\quad
    \text{s.t.}  \;{\text{\eqref{p1c},\eqref{p1d}}},\label{eqn_outer problem}
\end{equation}
and
\begingroup\makeatletter\def\f@size{9.2}\check@mathfonts
\def\maketag@@@#1{\hbox{\m@th\normalsize\normalfont#1}}%
\begin{align}
    \text{(P2-Inner)} &\; g^*(\mathbf{T}) =\max_{\mathbf{w}_0, \mathbf{w}_c}\log_2\left(\frac{\sigma^2 + |\mathbf{h}_1^H (\mathbf{T})\mathbf{w}_c |^2}{\sigma^2 + |\mathbf{h}_2^H (\mathbf{T})\mathbf{w}_c |^2}\right) \nonumber\\
    \text{s.t.} &\quad \min_{k=1, 2} \log_2\left(1 + \frac{|\mathbf{h}_k^H\left( \mathbf{T} \right) \mathbf{w}_0|^2 }{\sigma^2 + |\mathbf{h}_k^H\left( \mathbf{T} \right) \mathbf{w}_c |^2}\right) \geq r_{ms}, \nonumber\\
    & \quad |\mathbf{w}_0|^2 + |\mathbf{w}_c|^2 \leq P.\label{eqn_inner problem}
\end{align}\endgroup
Next, we employ the SDR to solve the inner problem, as it can yield a globally optimal solution to (P2-Inner).\vspace{-9pt}

\subsection{Inner-layer: Beamforming Optimization with Given $\mathbf{T}$}\label{innerProb}
Specifically, we let $\mathbf{Q}_0=\mathbf{w}_0\mathbf{w}_0^H\succeq {\bf{0}}$ and $\mathbf{Q}_c=\mathbf{w}_c\mathbf{w}_c^H\succeq {\bf{0}}$, with rank$(\mathbf{Q}_0)=\text{rank}(\mathbf{Q}_c)=1$. By further leveraging the monotonic property of the logarithm function, we can recast (P2-Inner) as
\begingroup\makeatletter\def\f@size{9.2}\check@mathfonts
\def\maketag@@@#1{\hbox{\m@th\normalsize\normalfont#1}}%
\begin{align}
    \text{(P3)} &\quad g^*(\mathbf{T}) =\max_{\mathbf{Q}_0\succeq {\bf{0}}, \mathbf{Q}_c\succeq {\bf{0}}}\frac{\sigma^2 + \mathbf{h}_1^H (\mathbf{T})\mathbf{Q}_c\mathbf{h}_1(\mathbf{T}) }{\sigma^2 + \mathbf{h}_2^H(\mathbf{T})\mathbf{Q}_c \mathbf{h}_2(\mathbf{T}) } \nonumber\\
    \text{s.t.} &\quad  \mathbf{h}_k^H(\mathbf{T})
    \mathbf{Q}_0 \mathbf{h}_k (\mathbf{T})- \tau_{ms}(\sigma^2 + \mathbf{h}_k^H(\mathbf{T}) \mathbf{Q}_c\mathbf{h}_k (\mathbf{T}) ) \geq 0, \nonumber\\
    &\quad k=1,2,\nonumber\\
    & \quad \text{Tr}(\mathbf{Q}_0 + \mathbf{Q}_c) \leq P,\; \text{rank}(\mathbf{Q}_0)=\text{rank}(\mathbf{Q}_c)=1.
\end{align}\endgroup

Although (P3) is still a non-convex optimization problem w.r.t. $\mathbf{Q}_0$ and $\mathbf{Q}_c$, we can relax the rank-one constraints in (P3) and introduce the Charnes-Cooper transformation, i.e.,
\begin{equation}\label{eqn_Charnes-Cooper}
\mathbf{Q}_c=\mathbf{Z}/\xi, \mathbf{Q}_0=\mathbf{\Gamma}/\xi, \xi \geq 0.
\end{equation}
Then, we can rewrite (P3) as
\begingroup\makeatletter\def\f@size{9.2}\check@mathfonts
\def\maketag@@@#1{\hbox{\m@th\normalsize\normalfont#1}}%
\begin{align}
     \max_{\mathbf{Z}\succeq {\bf{0}}, \mathbf{\Gamma}\succeq {\bf{0}}, \xi }&\;\xi + {\mathbf{h}_1^H( \mathbf{{T}}) \mathbf{Z} \mathbf{h}_1 (\mathbf{{T}})}/{\sigma^2} \nonumber\\
    \text{s.t.} \;&\xi + {\mathbf{h}_2^H( \mathbf{{T}}) \mathbf{Z} \mathbf{h}_2 (\mathbf{{T}})}/{\sigma^2} = 1, \nonumber\\
     &\mathbf{h}_k^H( \mathbf{{T}} ) \mathbf{\Gamma} \mathbf{h}_k (\mathbf{{T}}) \!-\! \tau_{ms}( \xi \sigma^2 + \mathbf{h}_k^H\left( \mathbf{{T}} \right) \mathbf{Z} \mathbf{h}_k (\mathbf{{T}})) \geq 0, \nonumber\\
     &k=1,2, \label{eqn_SDP}\\
     &\text{Tr}(\mathbf{Z}+\mathbf{\Gamma}) \leq P\xi.\nonumber
\end{align}\endgroup
Note that problem \eqref{eqn_SDP} is convex, which thus can be optimally solved through the interior-point algorithm.

Let $\mathbf{Z}^*$, $\mathbf{\Gamma}^*$ and $\xi^*$ denote the optimal solution to problem \eqref{eqn_SDP}. Then, it must hold that $\text{rank}^2(\mathbf{Z}^*)+\text{rank}^2(\mathbf{\Gamma}^*) \le 3$ for problem \eqref{eqn_SDP}\cite{W. Mei}, where the constant “3” denotes the total number of linear equalities and inequalities in problem \eqref{eqn_SDP}. As such, the optimal solutions to \eqref{eqn_SDP} satisfy $\text{rank}(\mathbf{Z}^*)=\text{rank}(\mbf{\Gamma}^*)=1$. Then, we can retrieve the optimal $\mathbf{Q}_0$ and $\mathbf{Q}_c$ based on \eqref{eqn_Charnes-Cooper} and then the optimal $\mathbf{w}_0$ and $\mathbf{w}_c$ via the singular value decomposition (SVD).\vspace{-6pt}
\begin{figure*}[t]
\centering
\subfloat[]{
    \includegraphics[width=0.26\linewidth]{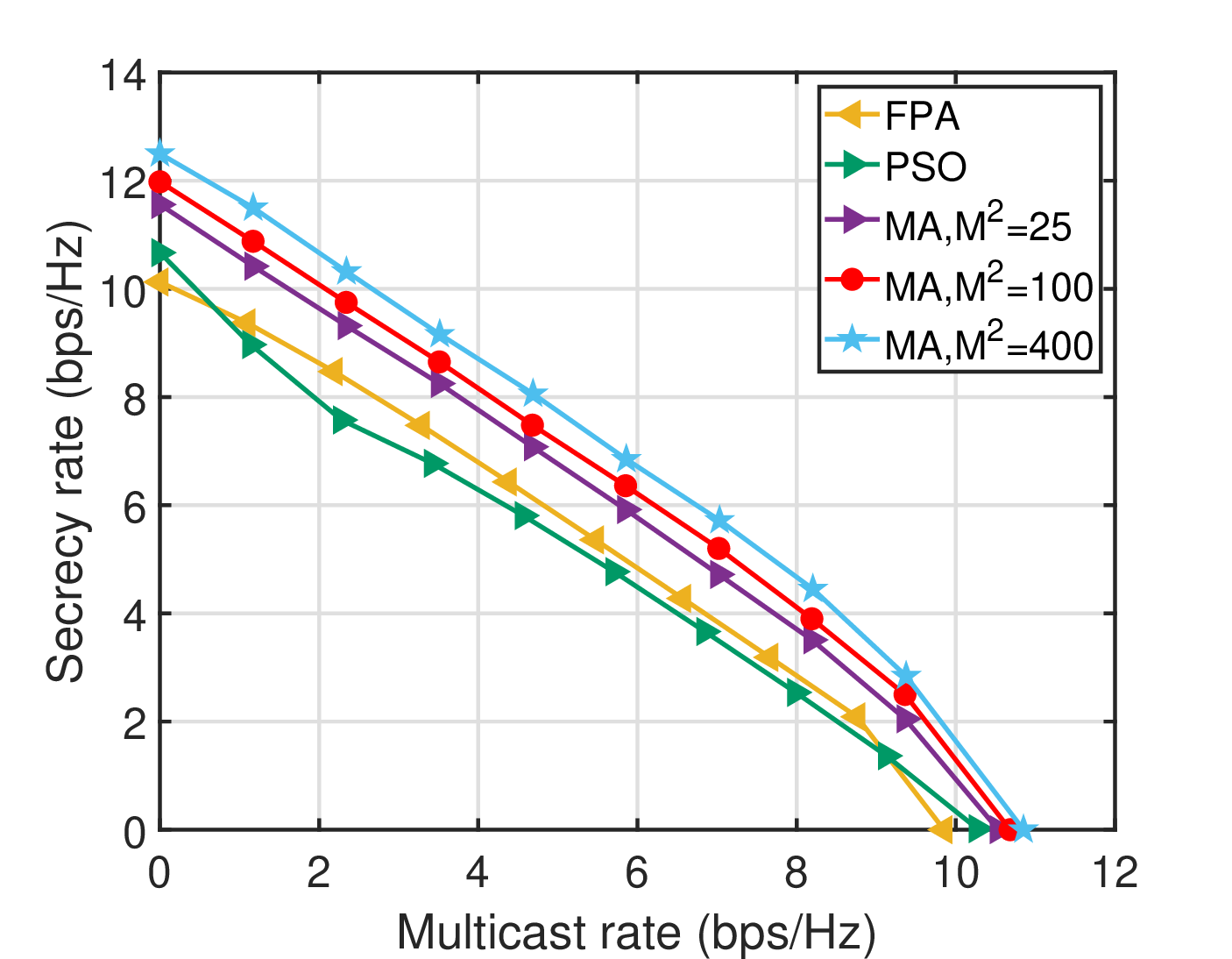}
    % \label{Fig_TxRegion}
}
\subfloat[]{
    \includegraphics[width=0.26\linewidth]{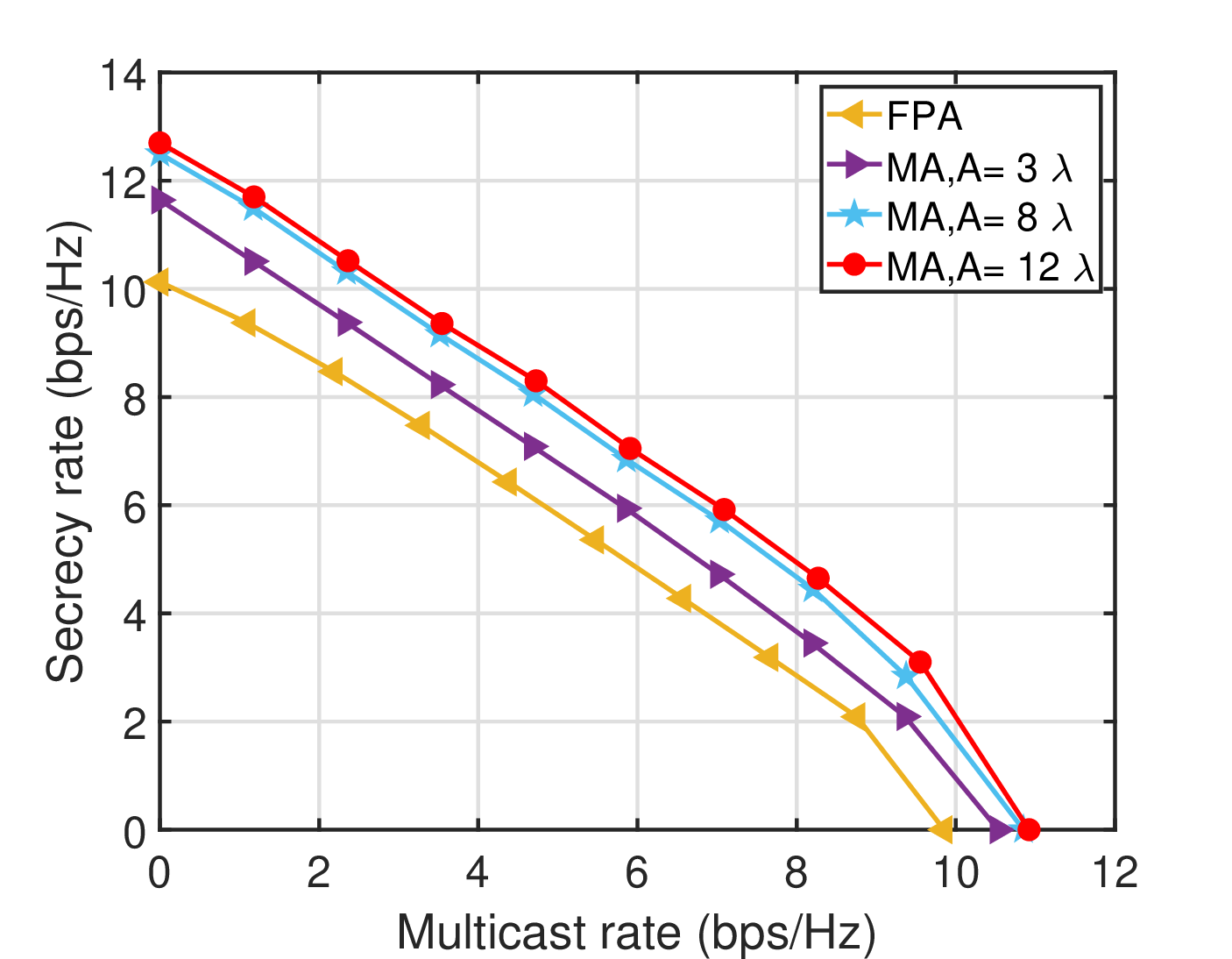}
    % \label{Fig_InfTemp}
}
\subfloat[]{
    \includegraphics[width=0.26\linewidth]{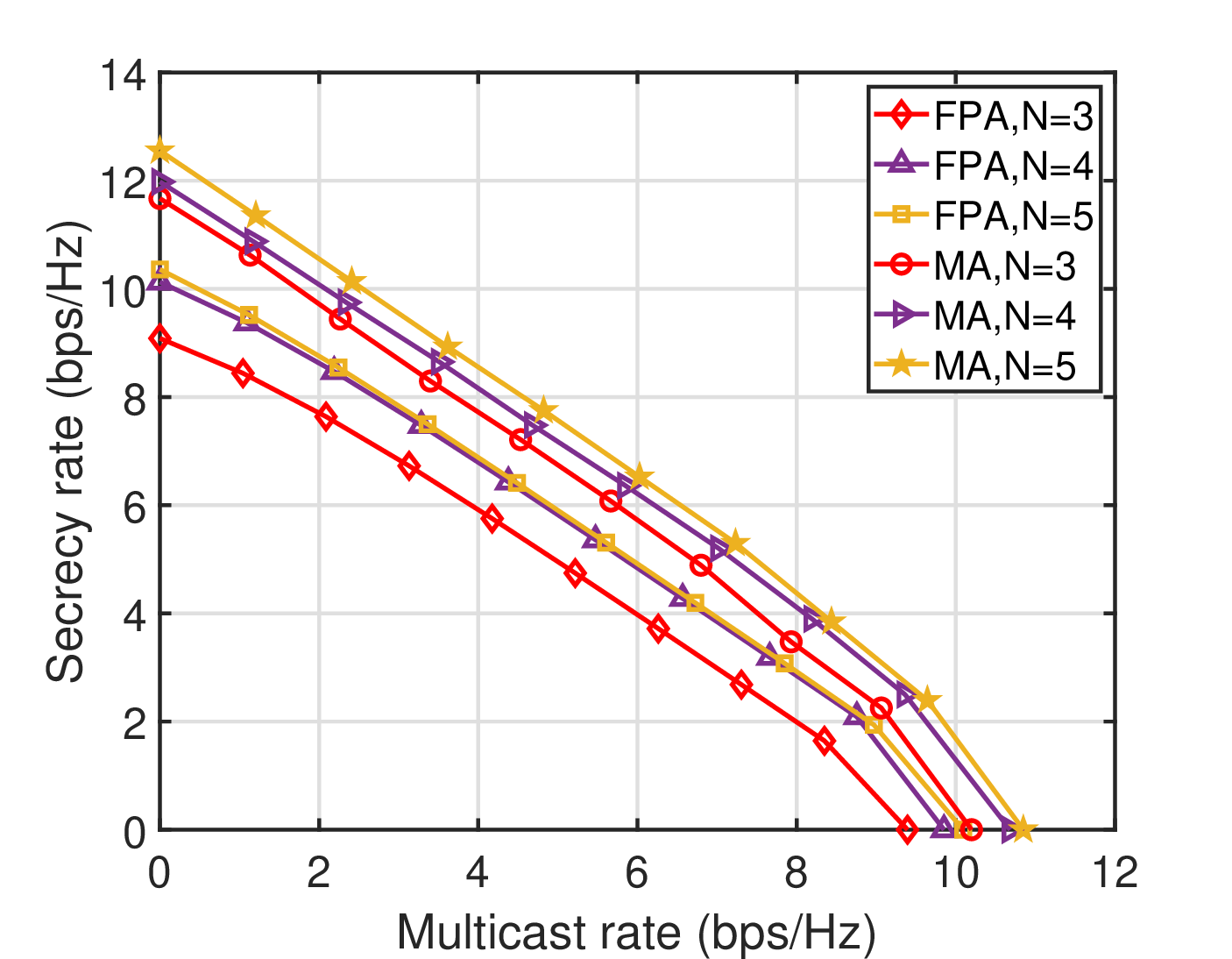}
    % \label{Fig_PathNums}
}
\caption{Achievable secrecy rate region versus (a) the number of sampling points per dimension; (b) the size of transmit region per dimension; and (c) the number of BS antennas}
\label{fig_sim}
\vspace{-19pt}
\end{figure*}

\subsection{Outer-layer: APV Optimization}
By employing the proposed algorithm in Section \ref{innerProb}, we can obtain the value of $g^*(\mathbf{T})$ for any given $\mathbf{T}$. However, due to the lack of analytical form of $g^*(\mathbf{T})$, it is still difficult to obtain the optimal $\mbf{T}$ that yields the maximum value of $g^*(\mathbf{T})$. To tackle this difficulty, we propose to find a high-quality suboptimal $\mbf{T}$ via discrete sampling as in \cite{W. Mei2024Movable}. Compared to other widely adopted antenna position optimization methods, e.g., gradient descent and PSO, the discrete sampling method is applicable to any problem structure, while ensuring local optimality via iterative refinement.

Specifically, we uniformly sample the horizontal/vertical
dimension of the transmit region $C_t$ into $M\;(M\gg N)$ discrete
points, with a spacing given by $\delta_s = A/M$. By this means, the continuous transmit region is discretized into $M^2$ sampling points, with the position of the $(i,j)$-th sampling point given by $\mathbf{p}_{i,j}=[-\frac{A}{2}+\frac{iA}{M},-\frac{A}{2}+\frac{jA}{M}]^T$, $i,j\in\mathcal{M}\triangleq\{1,2,\cdots,M\}$. Let $\mathcal{P}=\{\mathbf{p}_{i,j}|i,j\in\mathcal{M}\}$ denote the set of all sampling points. Next, we propose a sequential search method to optimize $\bf T$.

First, we construct a set of initial positions of MAs denoted by $\mathbf{\tilde{t}}_n\in\mathcal{P}, \forall n \in \mathcal{N}$. We consider that the position of the $n$-th MA, i.e., $\mathbf{\Tilde{t}}_n$, needs to be updated in the $n$-th iteration of the sequential search, with the positions of all other $(N-1)$ MAs being fixed. Let $\mathcal{P}_n$ denote the set of all feasible sampling points for the $n$-th MA, which is given by
\begin{equation}\label{eqn_DiscreteSet}
    \mathcal{P}_n=\{\mathbf{p}\in\mathcal{P}|\,\,||\mathbf{p}-\mathbf{\Tilde{t}}_m||_2\ge D_{\min},\forall m \in\mathcal{N},m\ne n\}.
\end{equation}

Then, in the $n$-th iteration, we solve the following optimization problem, 
\begin{equation}
    \text{(P2-Outer-$n$)} \quad \mathop {\max}\limits_{\mbf{t}_n \in {\cal P}_n} g^*(\mbf{T}),
\end{equation}
for which the optimal solution can be obtained via enumeration. Denote by $\hat{\mathbf{t}}^*_n$ the optimal solution to (P2-$n$). Next, we update $\tilde{\mathbf{t}}_n=\hat{\mathbf{t}}^*_n$ and proceed to solve (P2-Outer-$(n+1)$). It is evident that this sequential search yields non-decreasing values of $g^*(\mbf{T})$; hence, its convergence is ensured. By this means, we can find a locally optimal APV upon the convergence of the algorithm. It can be shown that the computational complexity of solving the inner problem is ${\cal O}(N^{6.5})$ via the SDR \cite{W. Mei}, which is also the complexity order for the conventional PHY-SI with FPAs and the time-sharing strategy. As solving the outer problem needs to invoke the SDR at most $NM^2$ times, the worst-case complexity of our proposed two-layer algorithm is given by ${\cal O}(N^{7.5}M^2)$. \vspace{-6pt}

\section{Numerical Results}
In this section, numerical results are provided to validate the performance of our proposed algorithm for MA-enhanced PHY-SI. Unless otherwise specified, the simulation settings are as follows. The operating frequency is set as 5 GHz. The minimum distance between any two adjacent MAs is set as $D_{\min}=\frac{\lambda}{2}$ with $\lambda=0.06$ meter (m). The distance between the BS and the two users is assumed to be 70 m. We employ the field-response channel model presented in \cite{L. Zhu2025tutorial} for MAs in the simulation. The number of transmit paths from the BS to each user is set as $L=7$. Let $\beta_{k,p}$ denote the path gain of the $p$-th transmit path from the BS to user $k$, which is assumed to follow the complex Gaussian distribution, i.e., $\beta_{k,p}\sim\mathcal{CN}(0,\rho d_k^{-2.8}/L),k = 1,2, 1 \le p \le L$, where $\rho=\left(\frac{\lambda}{4\pi}\right)^2$ represents the path loss at the reference distance of 1 m and ``$1/L$'' is due to the power allocation for all paths. The elevation and azimuth AoAs for all transmit paths are assumed to be independently and identically distributed (i.i.d.) variables following the uniform distribution within $[0,\pi]$. The BS's maximum transmit power is $P=41$ dBm, and the average noise power $\sigma^2$ is $-80$ dBm.

First, by varying the value of $r_{ms}$, Fig.\,\ref{fig_sim}(a) shows the secrecy rate regions by FPAs and MAs versus the number of sampling points per dimension $M$, with $N=4$. The PSO algorithm is also added as a benchmark. The size of transmit region per dimension is $A=8\lambda$. It is observed that 25 sampling MAs are sufficient for the MAs to yield a larger secrecy rate region compared to FPAs. Moreover, this rate region expands with $M$, as a larger number of sampling points improves the resolution of MA movement to create more favorable channel conditions. Nonetheless, the increase in the region size is observed to be small, implying that a low-to-moderate resolution may achieve near-optimal performance of continuous searching. Furthermore, it is observed that the PSO algorithm may even yield a worse performance than the FPA benchmark for certain multicast rates.

Next, Fig.\,\ref{fig_sim}(b) shows the secrecy rate regions by FPAs and MAs versus the size of transmit region per dimension, with $N=4$ and the sampling interval fixed at $0.4\lambda$. It is observed that all schemes employing MAs outperform FPAs, and their performance initially improves with increasing $A$ and eventually saturates. This is because a larger transmit region enables the MAs to harness more pronounced spatial diversity gain within a certain range.

Lastly, we plot the secrecy rate regions by FPAs and MAs versus the number of antennas $N$ in Fig.\,\ref{fig_sim}(c), with $M=20$ and $A=4\lambda$. It is observed that for both MAs and FPAs, the larger the number of antennas, the larger the secrecy rate region. This is because an increase in the number of antennas can offer a higher beamforming gain for both secrecy and multicast transmission. Besides, with the same number of antennas, the rate region by MAs is larger than that by FPAs. It is also observed from all of the above figures that the secrecy rate regions achieved by MAs can outperform the time-sharing strategy, unlike the single-MA case presented in Section \ref{singleMACase}.\vspace{-9pt}

\section{Conclusion}
In this letter, we studied a performance optimization problem for an MA-enhanced PHY-SI system, aiming to maximize the secrecy rate subject to the constraint on the multicast rate. Our theoretical analysis revealed the capability of MAs to cater to signal transmission of different types. A two-layer optimization framework was proposed to obtain a high-quality suboptimal solution to the SRM problem. Numerical results showed that MAs can greatly enhance the secrecy rate region for PHY-SI compared to FPAs. It is interesting to investigate more general MA-enhanced PHY-SI systems with multiple eavesdroppers or other integration models, and develop more efficient algorithms for rate region characterization in future.

\end{document}